# Quantum discord in the central spin model

V.E. Zobov*

L.V. Kirensky Institute of Physics, Russian Academy of Sciences, Siberian Branch, 660036, Krasnoyarsk, Russia,

**ABSTRACT**

The quantum properties of dynamic correlations in a system of an electron spin surrounded by nuclear spins under the conditions of free induction decay (FID) and spin echo have been studied. Analytical results for the time evolution of mutual information, classical part of correlations, and quantum part characterized by quantum discord have been obtained within the central spin model in the high-temperature approximation. The same formulas describe the quantum discord in both the FID and the spin echo although the forms of the dependences are different because of difference in the parameters entering into the formulas. Discord of the spin echo compared with the FID has a strong dependence on time at short times, and it tends to zero with decreasing of the magnetic field, whereas in the case of the FID it reaches a plateau.

## 1. INTRODUCTION

In recent years there has been grown interest in studying quantum information[1,2], quantum computers[2,3], and in the separation of correlations on quantum and classical parts[4]. For example, highly mixed state of qubits in the deterministic quantum computation with one qubit (DQC1)[5] is believed to perform a task exponentially faster than any classical algorithm. Even at high temperatures at which there is no quantum entanglement between the control qubit and the mixed ones, but there is a quantum discord[6] – the discrepancy between quantum versions of two classically equivalent expressions for mutual information[4]. What quantum properties are responsible for the DQC1 performance, has not yet been established[6-11]. In order to study physics of quantum discord we now consider a similar system – the central spin model[12-15], consisting of an electron spin coupled to $n$ nuclear spins. The sample is magnetized in a strong magnetic field. After the $90^0$ pulse of resonance magnetic field on the electron spin, the free induction decay signal (FID) $g_f(t)$ will be observed. If after a time $t$ the second $180^0$ resonance pulse acts on the electron spin, then the spin echo[16] with the amplitude $g_e(t)$ will be observed at the time 2t. The spin echo is a powerful method for studying the local properties of solids and liquids[16,17]. At the same time, this is an implemented experimentally example of the Loschmidt echo when studying nonequilibrium processes in multispin systems[18,19].

As examples, we can cite Nitrogen-Vacancy (NV) centers in diamond[13,14,20,21], impurities in solid silicon[12,22], quantum dots[15] etc..
1)      Diamond. Electron of a NV center surrounded by atoms (nuclei) of carbon. The main isotope $^{12}C$ is spinless, but the spin of rare (1%) isotope $^{13}C$ is half.
2)      Silicon. Electron on a phosphorus impurity surrounded by silicon nuclei. The isotope $^{28}Si$ has the zero spin, in the rare (about 5%) isotope $^{29}Si$ the spin is half.
We assume a low density of electron spins and ignore interactions between them. Flips of the nuclear spins cause the spin-echo decay. Possible causes are a spin-lattice interaction, a spin-spin interaction, a hyperfine interaction with the electron spin (back action). We will consider only the latter mechanism.

There are a huge number of theoretical and experimental works to the spin echo, including spin-echo attracted to studying classical and quantum properties of the reservoir of nuclear spins[19-21,23,24]. We do not know the works on the calculation of discord under the conditions of spin echo. In the work[22] on the phosphorus impurity in silicon, the authors experimentally prepared a special two-spin state (Bell-state) of the electron-nucleus system with nonzero discord. On the other hand, in a series of papers[6-10] the authors estimated the quantum discord at working DQC1 model to clarify the role of quantum correlations. For example, such working can be the calculation of the trace of a matrix. The real or imaginary part of the desired trace is found by measuring the average projections $\langle \hat{S}_x \rangle$ or $\langle \hat{S}_y \rangle$ on the *x* or *y*-axis, respectively. The working of DQC1 computer was demonstrated on simple systems, including $Ce^{3+}$ impurities in a $CaF_2$ crystal observed by ESR method[25]. We note similar papers on the calculation of the quantum discord[26-28]. In present work, we apply to the electron-nuclei spin echo the approach to the calculation of discord developed for the DQC1 model.

*rsa@iph.krasn.ru

## 2. THEORY

### 2.1 Free induction decay and spin echo

Our system of an electron spin and *n* nuclear spins in a strong static magnetic field we will described by the simple model Hamiltonian with one member of the hyperfine interaction

$$H = \omega_e S_z - \sum_j \omega_j I_{jz} + S_z \sum_j A_{jx} I_{jx}, \qquad (1)$$

where $\omega_e$ and $\omega_j$ are the Larmor frequencies of the electron, *S* = ½, and nuclear, *I* = ½, spins, respectively; $\hat{I}_{j\alpha}$ is α-component of the *j*th spin operator (α = *x*, *y*, *z*), $A_{j\alpha}$ is the hyperfine interaction constant. The Hamiltonian (1) with *x*-component of the hyperfine interaction and the field along the *z*-axis describe well, for example, paramagnetic centers in diamond and malonic acid. For phosphorus impurity in silicon, the Hamiltonian with *z*-component of the hyperfine interaction and the field along the *x*-axis

$$\hat{H} = \omega_e \hat{S}_z - \sum_j \omega_j \hat{I}_{jz} + \hat{S}_z \sum_j A_{jz} \hat{I}_{jz}$$

is needed. It is known that quantum discord does not change under a unitary transformation of one of two subsystems[4] (rotation of nuclear spins now). Both Hamiltonians give the same result. We choose the first one.

We obtain the central spin model in the high-temperature approximation, since the polarization for the electron spin in the magnetic field is small, $\beta_S = \hbar\omega_e / kT \approx 10^{-2} \ll 1$, and the polarization for the nuclear spins is three orders of magnitude smaller $\beta_I \ll \beta_S$. For this reason, the equilibrium density matrix is taken in the form

$$\hat{\rho}_{eq} = (1 - \beta_S \hat{S}_z)/Z, \qquad (2)$$

where $Z = 2^{n+1}$ is the partition function.

The density matrix describing the evolution of the state of the system can be written in the form

$$\hat{\rho}(t) = \frac{1}{Z}[1 - \frac{\beta_S}{2}\{\hat{S}_+ \hat{U}^+_{(f,e)}(t) + \hat{S}_- \hat{U}^-_{(f,e)}(t)\}] = \frac{1}{Z}[1 - \beta_S \Delta\hat{\rho}(t)], \qquad (3)$$

both in cases of the FID (*f*) and of the spin echo (*e*) with

$$\hat{U}^+_f(t)\hat{S}_+ = \exp(-it\hat{H})\hat{S}_+ \exp(it\hat{H}) = \exp(-it\hat{H}_+)\exp(it\hat{H}_-)\hat{S}_+, \qquad (4)$$

$$\hat{U}^+_e(t)\hat{S}_+ = \exp(-it\hat{H})\hat{P}_{180} \exp(-it\hat{H})\hat{S}_- \exp(it\hat{H})\hat{P}_{180} \exp(it\hat{H}) =$$
$$= \exp(-it\hat{H}_+)\exp(-it\hat{H}_-)\exp(it\hat{H}_+)\exp(it\hat{H}_-)\hat{S}_+ \qquad (5)$$

respectively, where $\hat{S}_\pm = \hat{S}_x \pm i\hat{S}_y$, and $\hat{P}_{180}$ is the rotation operator of the spin *S* by $180^0$ about the *x* axis. $\hat{H}_+$ and $\hat{H}_-$ are the Hamiltonians for fixed values of the projection of the electron spin $S_z = +1/2$ and $S_z = -1/2$. $\hat{U}^-_{(f,e)}$ is the operator that is Hermitian conjugate to $\hat{U}^+_{(f,e)}$.

When the $180^0$ pulse flips the electron spin, then it changes hyperfine interaction fields of the electron spin on nuclear spins (see Eq. (5)). Since the nuclear spins before and after the pulse rotate in the different fields, it does not fully compensate for the phase and the decay of the spin echo is observed. We obtained for signals the FID and for the echo amplitude[29]

$$g_{(f,e)}(t) = \langle \hat{S}_x(t) \rangle / \langle \hat{S}_x(0) \rangle = \operatorname{Re} Tr(\hat{U}^+_{(f,e)}(t)) = \prod_j (1 - v_j), \qquad (6)$$

where

$$v_j = v_{jf} = 2n_{jx}^2 \sin^2 \frac{t\Omega_j}{2} \text{ for FID (4), and } v_j = v_{je} = 8n_{jx}^2 n_{jz}^2 \sin^4 \frac{t\Omega_j}{2} \text{ for spin echo (5),} \qquad (7)$$

$$\Omega_j^2 = (\omega_j^2 + A_{jx}^2/4), \quad n_{jx} = A_{jx}/(2\Omega_j), \quad n_{jz} = \omega_j/\Omega_j.$$

To carry out the calculations of Eqs. (4) – (5) we used the property of exponential operators for $I = \frac{1}{2}$ given by

$$\exp(-it\Omega n_x \hat{I}_x - it\Omega n_z \hat{I}_z) = \cos(\Omega t/2) - i2\sin(\Omega t/2)\{n_x \hat{I}_x + n_z \hat{I}_z\}.$$

The signals (6) are expressed through the parameters $v_j$ in the same way for the FID and the spin echo. However, these parameters (7) themselves are different depending on the time and on the magnetic field as Figure 1 shows.

## 2.2 Quantum and classical correlations

It is convenient to carry out the next calculations in the orthogonal basis $|\Theta_k\rangle$ consisting of the $N = 2^n$ eigenfunctions of the evolution operator of the nuclear spin system[6], when we have

$$\hat{U}^\pm_{(f,e)} |\Theta_k\rangle = e^{\pm i\Theta_k} |\Theta_k\rangle. \qquad (8)$$

Since in our model there is no interaction between the nuclear spins, we obtained a product of nuclear spin contributions, $e^{i\Theta_k} = \prod_j e^{i\Theta_k^j}$. At different $k$ the phase $\Theta_k^j$ of spin $j$ takes one of two values

$$\Theta_k^j = \pm 2\arcsin\sqrt{v_j/2}, \qquad (9)$$

and the observed signals (6) of the echo or the FID can be written as a product $g_{(f,e)}(t) = \prod_j \cos\Theta_k^j$ of $\cos\Theta_k^j = 1 - v_j$. This phase $\Theta_k^j$ is expressed through the parameter $v_j$ (7) in the same way by Eq. (9) for the FID and the spin echo. However, these parameters are different functions of the time and of the magnetic field (see Fig.1).

The density matrix (3) in this representation has the form

$$\hat{\rho}(t) = \frac{1}{Z}\{1 - \beta_S \sum_k [\hat{S}_x \cos\Theta_k - \hat{S}_y \sin\Theta_k] \otimes |\Theta_k\rangle\langle\Theta_k|\}, \qquad (10)$$

where $\hat{\Pi}_k = |\Theta_k\rangle\langle\Theta_k|$ are projectors on the orthogonal basis of the nuclear system. The density matrix (3) has the same form as the matrix of quantum computation in the DQC1 model[5,6]. Only they have specially prepared the unitary operator $\hat{U}_n^+$ whose trace should be calculated to solve the task, but we have the evolution operator $\hat{U}^+_{(f,e)}$. Now we apply to the density matrix (10) the approach to calculation of the quantum discord developed for the DQC1 model[6-10].

The information-theoretic measure of correlations between the two systems is the mutual information,

$$I(\hat{\rho}) = S(\hat{\rho}_S) + S(\hat{\rho}_I) - S(\hat{\rho}), \qquad (11)$$

where reduced density matrices are

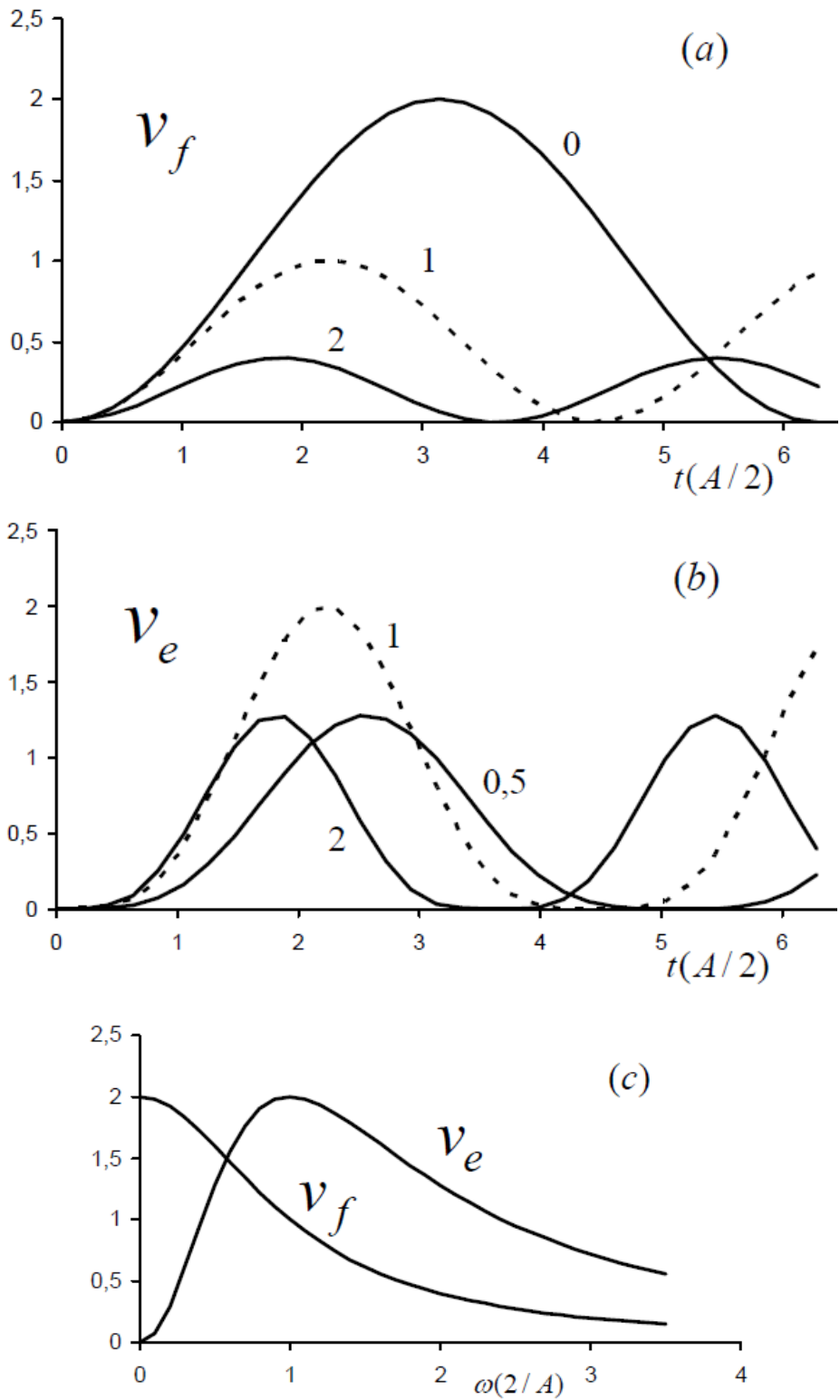

Figure 1. The parameters $v_f$ and $v_e$ as functions of the time at different values of the magnetic field (the number beside the curves are values $2\omega/A$) ((a) and (b)) and as functions of the magnetic field at $\sin^2(t\Omega/2)=1$ (c).

$$\hat{\rho}_I(t) = Tr_S \hat{\rho}(t) = 2\hat{E}_I/Z, \qquad \hat{\rho}_S(t) = Tr_I \hat{\rho}(t) = \frac{1}{2}[\hat{E}_S - \beta_S \hat{S}_x g(t)],$$

$\hat{E}_I$ and $\hat{E}_S$ are unit matrices. In the high-temperature approximation in the lowest (quadratic) polarization order we find:
for the von Neumann entropy

$$S(\hat{\rho}) = -Tr\{\hat{\rho}\log_2 \hat{\rho}\} = \log_2 Z - \frac{\beta_S^2}{2Z \ln 2} Tr(\Delta \hat{\rho})^2,$$

for the mutual information

$$I(\hat{\rho}) = \frac{\beta_S^2}{2\ln 2}\left\{\frac{1}{Z}Tr(\Delta\hat{\rho})^2 - \frac{1}{2}Tr_S(\Delta\hat{\rho}_S)^2 - \frac{1}{2^n}Tr_I(\Delta\hat{\rho}_I)^2\right\} = \frac{\beta_S^2}{8\ln 2}\left[1 - g^2(t)\right]. \tag{12}$$

$I(\hat{\rho})$ (12) is the measure of total correlations: quantum and classical. Following Neumann, to extract the classical part we should project the density matrix. For the electron spin $S = 1/2$ we have two projectors onto two states,

$$\hat{\Pi}_{S\pm} = \frac{1}{2} \pm (a_x \hat{S}_x + a_y \hat{S}_y + a_z \hat{S}_z), \tag{13}$$

with the quantization axis defined by the direction cosines $a_\alpha$ ($\alpha = x, y, z$). In our density matrix (10) is no z-component of the electron spin, so we take the cosines in the xy plane: $a_z = 0$, $a_x = \cos\varphi$, $a_y = \sin\varphi$.
After projecting, we have

$$\hat{\Pi}_S(\Delta\hat{\rho}(t)) = \frac{1}{2}\sum_k \left\{\hat{\Pi}_{S+}\cos(\varphi + \Theta_k)\otimes|\Theta_k\rangle\langle\Theta_k| - \hat{\Pi}_{S-}\cos(\varphi + \Theta_k)\otimes|\Theta_k\rangle\langle\Theta_k|\right\}.$$

After calculating the trace in the high-temperature approximation we find

$$I(\hat{\Pi}_S(\hat{\rho})) = \frac{\beta_S^2}{8\ln 2}\left\{\frac{1}{2^n}\sum_k \cos^2(\varphi + \Theta_k) - g^2(t)\cos^2\varphi\right\} = \frac{\beta_S^2}{16\ln 2}\left[1 - g^2(t) - K\cos 2\varphi\right], \tag{14}$$

where

$$K = \prod_j \cos^2 \Theta_k^j - \prod_j \cos 2\Theta_k^j = \prod_j (1-v_j)^2 - \prod_j [2(1-v_j)^2 - 1]. \tag{15}$$

Universal measure of classical correlations is obtained after calculating the maximum of the angle φ. The maximum of Eq. (14) achieved at φ=0 if K<0, and at φ=π/2, if K>0, and expressed in terms of the modulus of K,

$$\max_\varphi I(\hat{\Pi}_S(\hat{\rho})) = \frac{\beta_S^2}{16\ln 2}\left[1 - g^2(t) + |K|\right]. \tag{16}$$

Subtracting it from the total correlations, we find quantum discord - quantum part of correlations,

$$D = I(\hat{\rho}) - \max_\varphi I(\hat{\Pi}_S(\hat{\rho})). \tag{17}$$

We obtain the result for the discord,

$$D = \frac{\beta_S^2}{16\ln 2}\left[1 - g^2(t) - |K|\right], \tag{18}$$

and for its ratio to the total correlations,

$$\frac{D}{I(\rho)} = \frac{1}{2}\left[1 - \frac{|K|}{1 - g^2(t)}\right]. \tag{19}$$

## 3. DISCUSSION

Let us analyze the result. Initially, the magnetization is directed along the *x*-axis. With increasing time, the electron spin deviates. The different terms in the superposition state (10) deviate at different angles determined by the state of the nuclear spins. Small-time variation is small, $\sum_j v_j \ll 1$, the mutual information and the discord are small too,

$$I(\rho) = \frac{\beta_S^2}{4\ln 2}\sum_j v_j, \quad \frac{D}{I(\rho)} \approx \sum_j v_j. \tag{20}$$

If $v_j \ll 1$, but $\sum_j v_j > 1$, we have $\frac{D}{I(\rho)} \approx \frac{1}{2}[1 - \exp(-2\sum_j v_j)]$. At large times, the electron spin in the superposition (10) of different states is distributed more or less uniformly in the *xy* plane, and we have

$$I(\rho) = \frac{\beta_S^2}{8\ln 2}, \quad \frac{D}{I(\rho)} \approx \frac{1}{2}. \tag{21}$$

When measured with projection on any direction we lose half correlations. The authors[6] pointed out this circumstance when calculating the discord in DQC1 model.

Let us consider the interesting case of equal values $A_{jx} = A$ of the coupling constants of the electron spin with the *n* neighboring nuclear spins and equal Larmor frequencies $\omega_j = \omega$. Figure 2 shows the dependences of the correlation measures on the parameter *v* in this case.

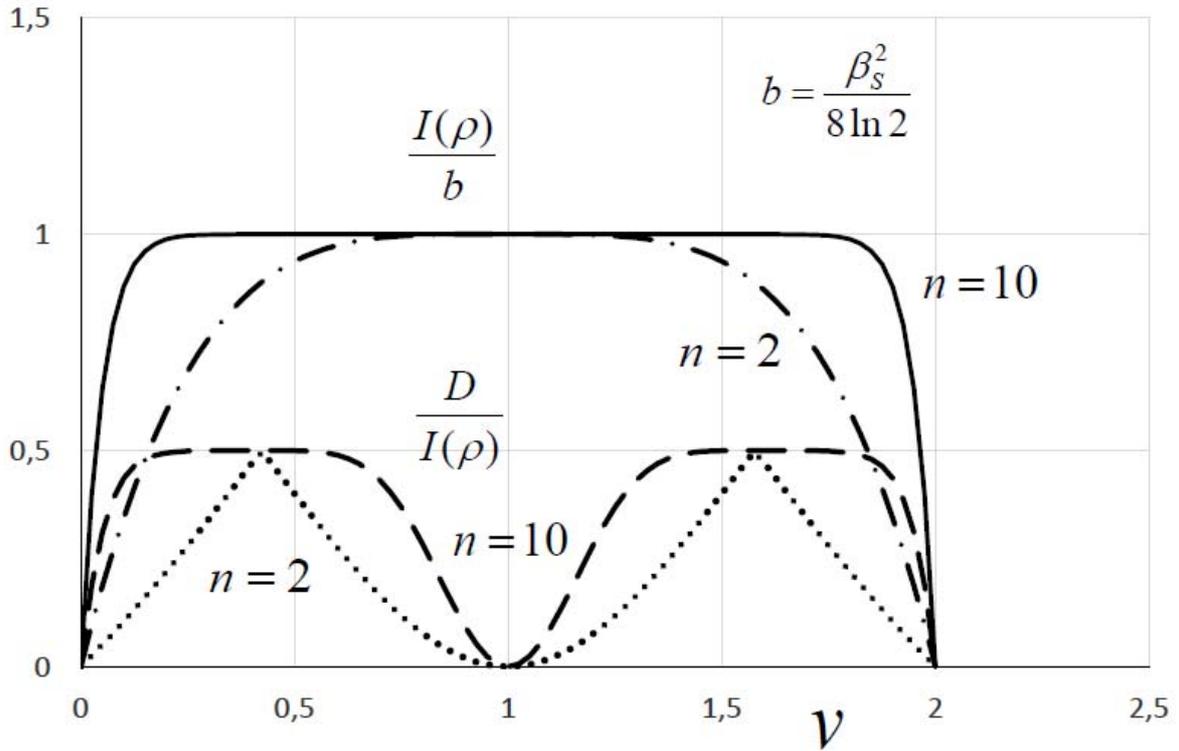

Figure 2. The mutual information and the quantum discord (its ratio to the total correlations) as functions of the parameter *v* for the systems of *n*=2 and n=10 nuclear spins equal coupling with the electron spin.

Universality depending on the parameter $v$ goes to different dependencies on the time and on the magnetic field for the FID and the spin echo, due to differences in the dependences of the parameters $v$ (7) on these quantities (see fig.1).

It is interesting on Fig.2 that when $v = 1$ the discord equals to zero, whereas the mutual information is maximal. Under these conditions we have $\Theta_k^j = \pm \frac{\pi}{2}$, and the electron spin in the superposition state (10) groups along one axis:

$y$-axis for odd $n$ ($\Theta_k = \sum_k \Theta_k^j = \pm \frac{\pi}{2} + 2\pi m_k$),

$x$-axis for even $n$ ($\Theta_k = \sum_k \Theta_k^j = \pm \pi + 2\pi m_k$).

When projected onto this axis, we perform measurements without loss, so the discord is zero.

Can we consider the state of completely the classic? For DQC1 model in similar circumstances authors[8] pay attention to the preservation of advantages over classical computer in the case $D=0$. We conclude that discord did not fully take into account the quantum properties such as quantum superposition and interference, which play an important role in performance DQC1 model of a quantum computer.

To understand, let us return to the measurement procedure. When in different states of the superposition (10) the electron spin is uniform distributed in the $xy$ plane, half the correlations are lost at projection (13) on any direction, and in Eq. (14) $2^{-n} \sum_k \cos^2(\varphi + \Theta_k) \approx 1/2$. The result will change radically if we first measure the state of the nuclear system, and then measure the state of the electron spin with the projector to the known correct axis. That is, when we take each measurement angle $\varphi_k = -\Theta_k$. Then in Eq. (14) we have $2^{-n} \sum_k \cos^2(\varphi_k + \Theta_k) = 1$, and we extract all the information without any loss. On a similar phenomenon, "unlocking" classical information mentioned in the work[30].

Thereby, when measuring $\langle \hat{S}_x \rangle$ or $\langle \hat{S}_y \rangle$ in the quantum system, we find the trace of the evolution by a single measurement. There is a summation on the spin $S$ of quantum information unknown to the observer. The summation of classical information known to observer requires performing sequentially the $N = 2^n$ operations of summation of the matrix elements on a classical computer. It is as for the spin echo and for the DQC1 computer.

**REFERENCES**

[1] Preskill, J., [Quantum Information and Computation], Lecture notes for physics 229, California Institute of Technology, Pasadena (1998); Vol. 1, Regular and Chaotic Dynamics, Moscow (2008).
[2] Nielsen, M. A. and Chuang, I. L., [Quantum Computation and Quantum Information], Cambridge University Press, Cambridge (2000); Mir, Moscow (2006).
[3] Valiev, K. A. and Kokin, A. A., [Quantum Computers:Expectations and Reality], Regular and Chaotic Dynamics, Moscow-Izhevsk (2001) [in Russian].
[4] Modi, K., Broduch, A., Cable, H, et al., "The classical-quantum boundary for correlations: discord and related measures," Rev. Mod. Phys. 84(4), 1655-1707 (2012).
[5] Knill, E. and Laflamme, R., "Power of one bit of quantum information," Phys. Rev. Lett. 81, 5672 (1998).
[6] Datta, A., Shaji, A. and Caves, C. M., "Qantum discord and the power of one qubit," Phys. Rev. Lett. 100, 050502 (2008).
[7] Wu, S., Poulsen, U. V. and Molmer K., "Correlations in local measurements on a quantum state, and complementarity as an explanation of nonclassicality," Phys. Rev. A 80, 032319 (2009).
[8] Dakic, B., Vedral, V. and Brukner, C., "Necessary and sufficient condition for nonzero quantum discord," Phys. Rev. Lett. 105, 190502 (2010).

[9] Datta, A. and Shaji, A., "Quantum Discord and Quantum Computing - An Appraisal," Int. J. Quantum. Inform. 9, 1787 (2011).

[10] Passante, G., Moussa, O. and Laflamme, R, "Measuring geometric quantum discord using one bit of quantum information," Phys. Rev. A 85, 032325 (2012).

[11] Morimae, T., Fujii, K. and Fitzsimons J. F. "On the hardness of classically simulating the one clean qubit model," Phys. Rev. Lett. 112, 130502 (2014).

[12] Witzel, W. M., Carroll, M. S., Cywinski, L. and Sarma, S. D., "Quantum decoherence of the central spin in a sparse system of dipolar coupled spins," Phys. Rev. B 86, 035452 (2012).

[13] Zhao, N., Ho, Sai-Wah, and Liu, Ren-Bao, "Decoherence and dynamical decoupling control of nitrogen vacancy centerelectron spins in nuclear spin baths," Phys. Rev. B 85, 115303 (2012).

[14] Hall, L. T., Cole, J. H. and Hollenberg, L. C. L., "Analytic solutions to the central spin problem for NV centres in diamond," arXiv:1309.5921.

[15] Hackmann, J. and Anders, F. B., "Spin noise in the anisotropic central spin model," Phys. Rev. B 89, 045317 (2014).

[16] Salikhov, K. M., Semenov, A. G. and Tsvetkov, Yu. D., [Electron Spin Echo and Its Applications], Nauka, Novosibirsk, (1976) [in Russian].

[17] Blumich, B., "Essential NMR for scientists and engineers," Springer-Verlag, Berlin, (2005); 'Technosfera', Moscow, (2007).

[18] Waugh, J. S., [New NMR methods in solid state physics], Cambridge University Press, Cambridge, (1976); Mir, Moscow, (1978).

[19] Jalabert, R. A. and Pastawski, H. M., "Environment-Independent Decoherence Rate in Classically Chaotic Systems," Phys. Rev. Lett. 86, 2490 (2001).

[20] Reinhard, F., Shi, F., Zhao, N., et al., "Tuning a Spin Bath through the Quantum-Classical Transition," Phys. Rev. Lett. 108, 200402 (2012).

[21] Laraoui, A., Dolde, F., Burk, C., et al., "High-Resolution Correlation Spectroscopy of $^{13}$C Spins Near a Nitrogen-Vacancy Center in Diamond," Nature Comm. 4, 1651 (2013).

[22] Rong, X., Jin, F. and Wang, Z., "Experimental protection and revival of quantum correlation in open solid systems," Phys. Rev. B 88, 054419 (2013).

[23] Zurek, W. H., "Decoherence, einselection, and the quantum origins of the classical," Rev. Mod. Phys. 75(3), 715-775 (2003).

[24] Fink, T. and Bluhm, H., "Distinguishing Quantum and Classical Baths via Correlation Measurements," arXiv:1402.0235.

[25] Mehring, M. and Mende, J., "Spin-bus concept of spin quantum computing," Phys. Rev. A 73, 052303 (2006).

[26] Kuznetsova, E. I. and Zenchuk, A. I., "Quantum discord versus second-order MQ NMR coherence intensity in dimmers," Phys. Lett. A 376, 1029 (2012).

[27] Zobov, V. E., "Quantum and classical correlations in high-temperature dynamics of two coupled large spins," Theor. Math. Phys. 177(1), 1377–1389 (2013).

[28] Chernyavskiy, A. Y., Doronin, S. I., and Fel'dman, E. B., "Bipartite quantum discord in a multiqubit spin chain," Phys. Scr. T160, 014007 (2014).

[29] Zobov, V. E., "Quantum and classical correlations in electron-nuclear shin echo," JETP, 119(5), 817-827 (2014).

[30] DiVincenzo, D. P., Horodecki, M., Leung, D. W., et al. "Locking Classical Correlations in Quantum States," Phys. Rev. Lett. 92, 067902 (2004).